\documentclass[showpacs,preprintnumbers,amsmath,amssymb,12pt,floatfix,epsfig]{revtex4}
\usepackage[usenames, dvipsnames]{color}
\usepackage{amssymb}
\usepackage{bm}
\usepackage{graphicx}
\usepackage{revsymb}
\usepackage{amsmath}
\usepackage{dcolumn}

\usepackage[normalem]{ulem}
\usepackage{longtable}
\usepackage{times}

\renewcommand\sout{\bgroup \color{red} \ULdepth=-.5ex \ULset}

\newcommand{\be}{\begin{equation}}
\newcommand{\ee}{\end{equation}}
\newcommand{\bea}{\begin{eqnarray}}
\newcommand{\eea}{\end{eqnarray}}

\renewcommand{\rm}[1]{\textrm{#1}}

\begin{document}
\title{EoS from terrestrial experiments:
 static and dynamic polarizations of nuclear density
}
\author{H. Sagawa\footnote{corresponding auther}
}
\address{RIKEN, Nishina Center, Wako 351-0198, Japan}
\address{Center for Mathematics and Physics, University of Aizu, Aizu-Wakamatsu, Fukushima 965-8580, Japan}

\author{S. Yoshida }
\address{Science Research Center, Hosei University, 2-17-1, Fujimi, Chiyoda, Tokyo 102-8160, Japan}

\author{Li-Gang Cao}

\address{School of Mathematics and Physics, North China Electric Power University, Beijing 102206, China
}

\begin{abstract}
We critically examine nuclear matter and neutron matter equation of state (EoS) parameters by using best available terrestrial experimental results.   The nuclear incompression modulus  $K_{\infty}$ is re-examined 
in comparisons with RPA results of modern relativistic and non-relativistic EDF and up-to-date experimental 
data of isoscalar giant monopole resonance energy of $^{208}$Pb.  The symmetry energy expansion coefficients $J$, $L$ and $K_{sym}$ are examined by recent FRDM mass model and the neutron skin of 
$^{48}$Ca extracted from   $(p,p')$ experiments.  
\end{abstract}
\maketitle
\section{Nuclear Equation of State (EoS)}
\begin{figure}[t]
\centering
\includegraphics[width=12cm,bb=0 0 720 540,clip]{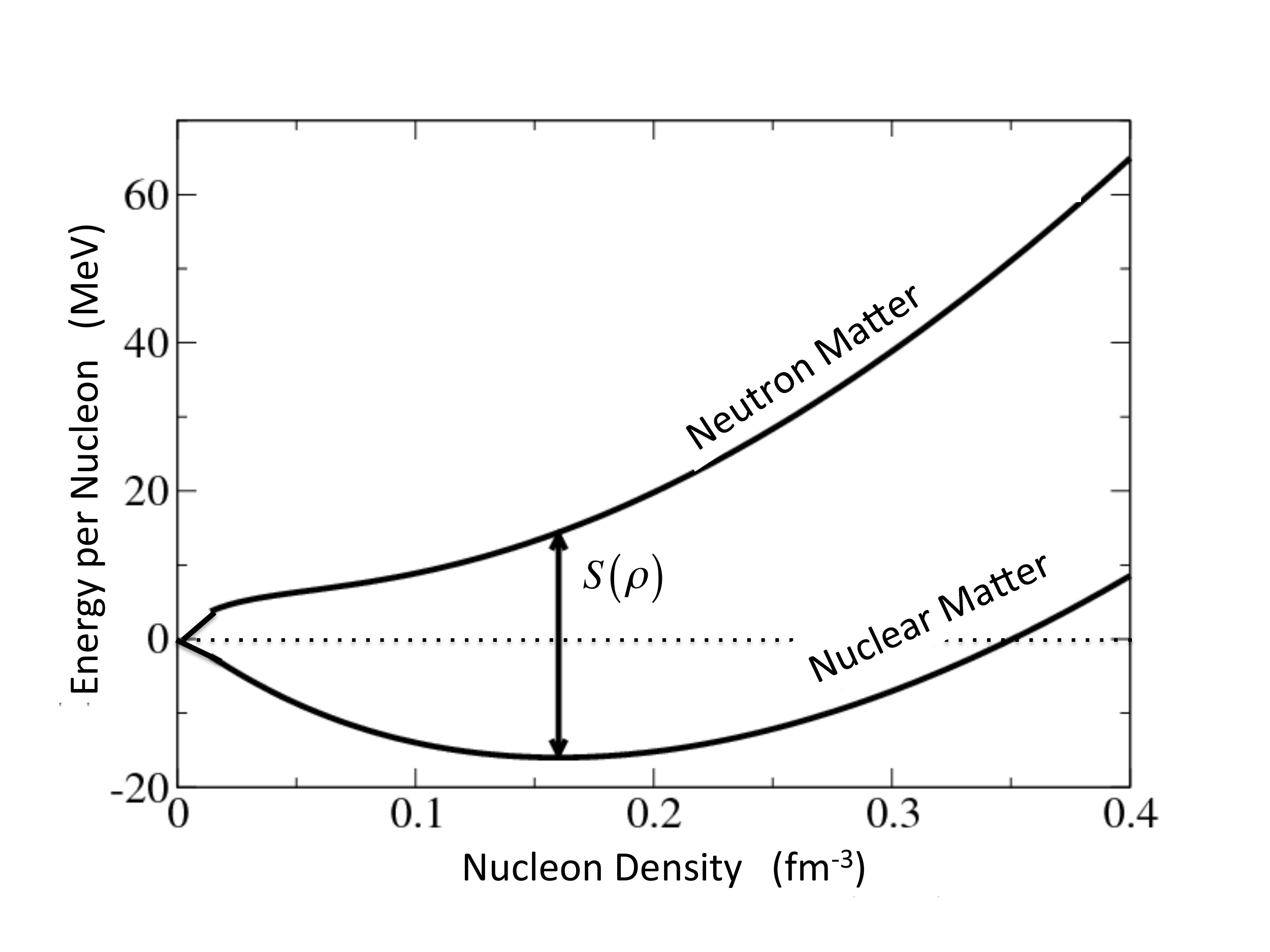}
\caption{Nuclear and neutron matter equation of state (EoS) near the saturation density. $S(\rho)$ is the symmetry energy in asymmetric nuclear matter.  \label{fig:EoS}}
\end{figure}
Contemporary nuclear science aims to understand the properties of strongly interacting bulk matter at the nuclear, hadronic and quark levels.   In addition to their intrinsic interest in fundamental physics, such studies have enormous impact on astrophysics, from the evolution of the early universe to neutron star structure. For example, a precise knowledge of the equation of state (EoS)  of neutron matter is essential to understand the physics of neutron stars and binary mergers, also predicted to be strong sources of gravitational waves.  On 17 August 2017,  
the LIGO and Virgo detectors    observed a gravitational wave  which  was produced by the last minutes of two neutron stars spiraling closer to each other and finally merging.  This gravitational wave is  named GW170817.

Although the size difference between the nucleus and the neutron star is almost 10$^{20}$ times,   there are deep and intimate relations between the two objects through nuclear matter and neutron matter EoS.  The EoS  of symmetric nuclear matter consisting of equal amount of neutrons and protons has been determined over a wide range of densities by terrestrial experiments.  As we can seen in Fig. \ref{fig:EoS}, the neutron matter EoS depends entirely the symmetry energy 
$S(\rho)$ on top of the symmetric nuclear matter. 
The nuclear symmetry energy characterizes the variation of the binding energy as the neutron to proton ratio of a nuclear system is varied. In other words,  
the symmetry energy constrains the force which determines the asymmetry between proton and neutron numbers  in a nuclear system. 
 It reduces the nuclear binding energy in nuclei and is critical for understanding properties of nuclei including the existence of rare isotopes with extreme proton to neutron ratios. 
More precisely, 
  its slope at saturation density shows a strong correlation with the neutron skin size of nuclei, and also gives the dominant baryonic contribution to the pressure in neutron stars.
  
Let us study hereafter  the EoS more quantitatively. The energy density of asymmetric nuclear matter can be expanded as 
\begin{equation}
\frac{E(\rho,\delta)}{\rho}\approx \frac{E_0(\rho,\delta=0)}{\rho}+E_{sym}(\rho)\delta^2, 
\label{eq:ED}
\end{equation}
where $\delta=(\rho_n-\rho_p)/\rho$ is the asymmetric coefficient and $\rho_n, \rho_p$ and $\rho=\rho_n+\rho_p$ are neutron, proton and total densities, respectively.  The symmetry energy is density by,  
\begin{eqnarray}
E_{sym}(\rho)=\frac{1}{2} \left. \frac{\partial^2 (E/A)}{\partial \delta^2} \right|_{\delta=0},    
\end{eqnarray}
is the symmetry energy. The pressure of nuclear matter at zero temperature is defined by
\begin{equation}
P(\rho)=\rho^2\frac{\partial(E(\rho)/\rho)}{\partial\rho}.
\label{eq:Pressure}
\end{equation}
At the saturation point,  $P(\rho_0$)=0,   
the EoS around the nuclear saturation density is essentilly determined by the incompressibility and the symmetry energy.  The energy density in  
symmetric matter is expressed by the Taylor  expansion around the 
saturation density $\rho_0$ as 
\begin{equation}
\frac{E_0(\rho,\delta=0)}{\rho}=\frac{E_0(\rho_0,\delta=0)}{\rho_0}+\frac{1}{2}(\frac{\rho-\rho_0}{3\rho_0})^2K_{\infty}+\frac{1}{3!}(\frac{\rho-\rho_0}{3\rho_0})^3Q+\cdot \cdot \cdot, 
\end{equation}
where $K_{\infty}$ is 
the incompressibility of nuclear matter and $Q$ is the skewness parameter.   $K_{\infty}$ is defined as the second derivative of the binding energy per particle with respect to the density at the saturation point
\begin{equation}
K_{\infty}= 9\rho^2\frac{\partial ^2(E_0/\rho)}{\partial \rho^2}\Bigr|_{\rho=\rho_0}, 
\label{eq:K}
\end{equation}
and  $Q$ is defined by
\begin{equation}
Q=27\rho^3\frac{\partial ^3(E_0/\rho)}{\partial \rho^3}\Bigr|_{\rho=\rho_0}.  
\label{eq:Q}
\end{equation}
The symmetry energy $E_{sym}$  in Eq. (\ref{eq:ED})  
is further expanded around the saturation density $\rho_0$ as
\begin{equation}
E_{sym}(\rho)\equiv S(\rho)=J+L\frac{(\rho-\rho_0)}{3\rho_0}+\frac{1}{2}K_{sym}\frac{(\rho-\rho_0)^2}{9\rho_0^2}
+\frac{1}{3!}(\frac{\rho-\rho_0}{3\rho_0})^3Q_{sym}+\cdot \cdot \cdot, 
\label{eq:sym}
\end{equation}
where
\begin{eqnarray}
J&=&S(\rho_0),    \\
L&=&3\rho_0\frac{\partial S(\rho)}{\partial \rho}\Bigr|_{\rho=\rho_0},   \\
K_{sym}&=&9\rho_0^2\frac{\partial^2 S(\rho)}{\partial \rho^2}\Bigr|_{\rho=\rho_0} ,  \\
Q_{sym}&=&27\rho_0^3\frac{\partial^3 S(\rho)}{\partial \rho^3}\Bigr|_{\rho=\rho_0}.  
\end{eqnarray}
Since neutron star contains a low fraction of protons,  the inner crust as well as global neutron star properties are sensitive to the symmetry energy parameters $J$ and $L$.  One can see easily the importance of the symmetry energy when one calculate the pressure of neutron matter at the saturation density,
\begin{eqnarray}\label{Pressure}
P(\rho_0)=\rho^2\frac{\partial(E/A)}{\partial \rho} \Bigr|_{\rho=\rho_0}=\frac{\rho_0}{3}L.
\end{eqnarray}

\begin{figure}[t]
\includegraphics[clip,width=12cm,bb=0 0 720 540]{
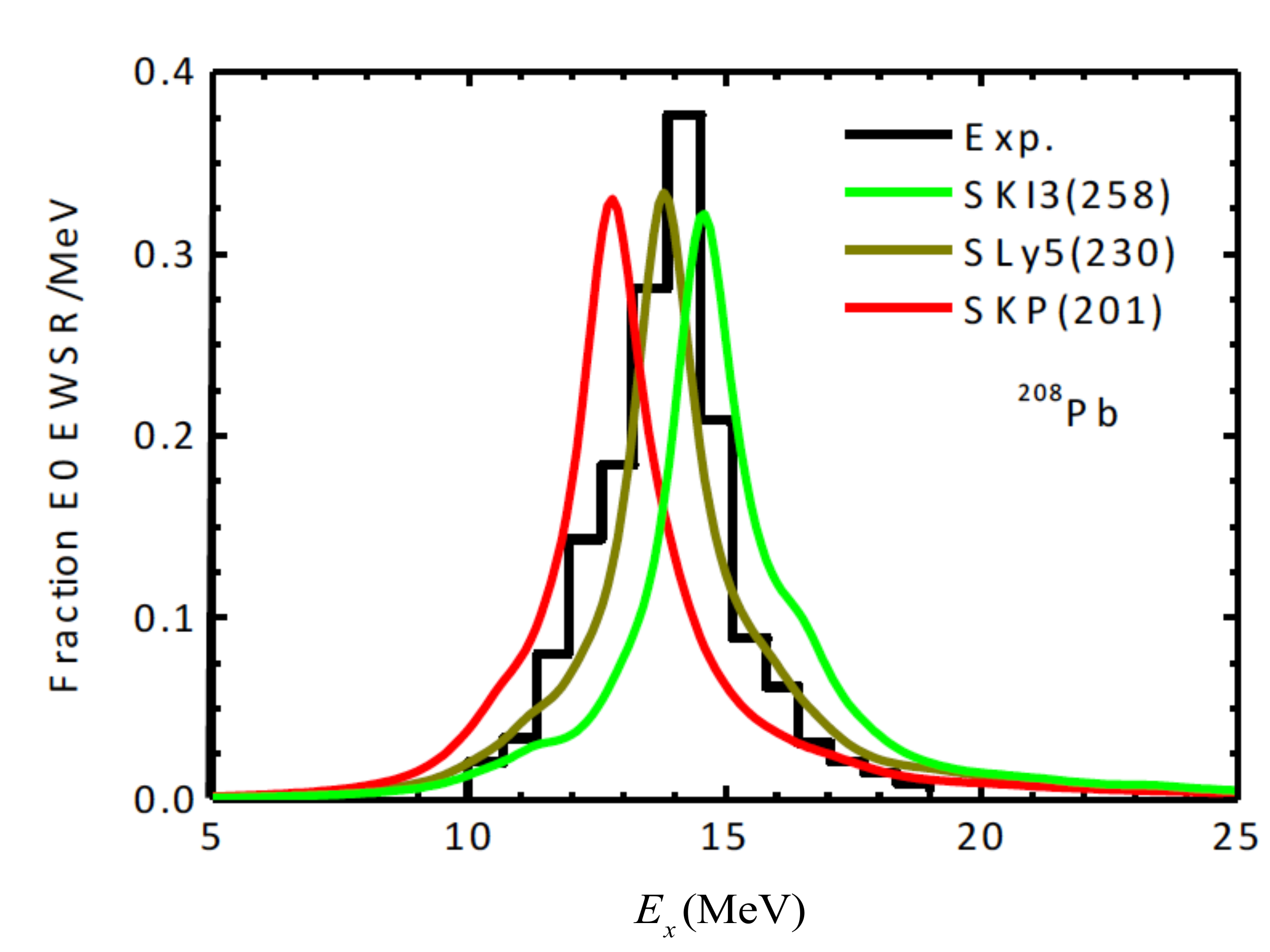}
\caption{The strength distributions  of ISGMR in $^{208}$Pb as a function of the excitation energy.  The theoretical results are obtained by self-consistent HF+RPA calculations with SkP, SLy5 and SkI3 interactions with K$_{\infty}$=201, 230 and 258MeV, 
respectively.  The experimental data is taken at  Texas A and M University, Cyclotron laboratory,
D. H. Youngblood et al., Phys. Rev. Lett. 82, 691 (1999) and 
 Y. -W. Lui, private communications.
\label{fig:Pb-ISGMR}}
\end{figure}

\section{GMR and nuclear incompressibility}\label{sec1} 

We study the incompressibility in relation with  terrestrial experiments.  
The incompressibility in finite nuclei has an analytic relation with the excitation energy of isoscalar giant monopole resonance (ISGMR) as 
\begin{equation}
\hbar \omega =\sqrt{\frac{\hbar^2 K_A}{m<r>^2}}, 
\label{eq:Kmono}
\end{equation}
where $m$ is a nucleon mass and $<r>^2$ is the mean square nuclear radius. Intuitively, this relation tells 
how the ISGMR,  so called breathing mode, can be affected by the solidness of nucleus.  If the ISGMR is a sharp single peak,
Eq. (\ref{eq:Kmono}) provides a precise empirical information of incompressibility in finite nuclei.
For the study of celestial observables such as supernovae or neutron stars, we need the information of nuclear matter incompressibility .    The incompressibility $K_A$ in finite nuclei may have
 contributions from the surface, the symmetry energy,  and the Coulomb 
energy on top of the nuclear matter incompressibility  as  an analogy of  the mass formula.   
 The relation can be   written as 
\begin{equation}
K_A=K_{\infty}+K_{surf}A^{-1/3}+K_{sym}\delta^2+K_{Coul}\frac{Z^2}{A^{4/3}}, 
\label{eq:KA}
\end{equation}
where $\delta=(N-Z)/A$.

\begin{table}
\caption{The centroid energies of ISGMR of $^{208}$Pb calculated by using relativistic and non-relativistic effective interactions. The  ISGMR energies are calculated by using the sum rule vales   $E^{cen}_{GMR}=m_1/m_0$ and E$^{con}_{GMR}=\sqrt{m_1/m_{-1}}$.  The theoretical results are obtained in the interval between 10.5 and 20.5 MeV. }
\begin{tabular}{ccccccccc}
\hline\hline
        & non-relativistic & & & relativistic  &    \\\hline
 Para. & K$_{\infty}$(MeV)         & E$_{GMR}$(MeV)  & E$^{con}_{GMR}$(MeV)&Para. & K$_{\infty}$(MeV)  & E$^{cen}_{GMR}$(MeV) & E$^{con}_{GMR}$(MeV) \\

\hline
 SKP      & 201 & 12.78 & 12.74  & NL1   & 211   &  12.59  &    12.57        \\
 SGII     & 215 & 13.48 & 13.44  & NLE   & 221   &  12.89  &    12.87     \\
 SKM$^*$  & 217 & 13.38 & 13.35  & NLC   & 224   &  13.42  &    13.37     \\
 SLy5     & 230 & 13.80 & 13.76  & FSU   & 230   &  14.27  &    14.23      \\
 SKI2     & 241 & 14.12 & 14.08  & IUFSU & 230   &  13.89  &    13.87      \\
 SK255    & 255 & 14.47 & 14.44  & NLBA  & 248   &  14.41  &    14.39      \\
 SKI3     & 258 & 14.63 & 14.59  & NL3   & 271   &  14.22  &    14.18      \\
 SGI      & 262 & 14.78 & 14.73  & TM1   & 281   &  15.14  &    15.07      \\
 SKA      & 263 & 14.62 & 14.57  & PK1   & 283   &  14.28  &    14.13      \\
 SKB      & 263 & 14.86 & 14.83  & NLSH  & 355   &  16.86  &    16.77      \\
 SKx      & 271 & 15.14 & 15.08  &    &    &        &     \\
 SIV      & 325 & 15.45 & 15.38  &    &    &        &     \\
 Z        & 330 & 16.75 & 16.67  &    &    &        &     \\
 E        & 333 & 16.78 & 16.70  &    &    &        &     \\
 SII      & 341 & 16.48 & 16.42  &    &    &        &     \\
 SIII     & 355 & 16.98 & 16.91  &    &    &       &      \\
  \hline\hline
\end{tabular}
\end{table}

\begin{table}
\caption{The ISGMR experimental data of $^{208}$Pb. }
\begin{tabular}{ccccccccc}
\hline\hline
               & UND\cite{1}    & Orsay \cite{2}   & J\"ulich \cite{3}& RCNP-U \cite{4}& RCNP-P\cite{5} & TAMU \cite{6}    & IUCF\cite{7}   & KVI \cite{8} \\
\hline
E (MeV)        & 13.6 $\pm$ 0.1 & 13.5 $\pm$ 0.3  & 13.8 $\pm$ 0.3 & 13.5 $\pm$ 0.2 &  13.7 $\pm$ 0.1 & 13.96 $\pm$ 0.2  & 13.9 $\pm$ 0.4 & 13.9 $\pm$ 0.3 \\
$\Gamma$(MeV)  &  3.1 $\pm$ 0.4 &  2.8 $\pm$ 0.2  &  2.6 $\pm$ 0.3 &  4.2 $\pm$ 0.3 &   3.3 $\pm$ 0.2 &  2.88 $\pm$ 0.2  &  3.2 $\pm$ 0.4 &  2.5 $\pm$ 0.4 \\
EWSR  (\%)        & 147 $\pm$ 18   &  307 $\pm$ 60   &           &   58 $\pm$ 3   &           &    99 $\pm$ 15   &  100 $\pm$ 20  &  110 $\pm$ 22  \\
\hline\hline
\end{tabular}
\end{table}

Experimental data of ISGMR have been obtained by inelastic scatterings of isoscalar probes, especially by $(\alpha,\alpha')$ inelastic scatterings.  The experimental cross sections are analyzed  by multipole decomposition analysis (MDA) to separate  
the monopole components from other multipoles with $L>0$.  
This method is very promising 
 since the cross sections with $L=0$  have peaks at the forward angle $\theta \sim 0^{\circ}$ and   other multipoles have peaks at larger angles $\theta > 0^0$. 
The MDA  technique  was thus applied  to extract 
the strength distributions  of IS monopole (GMR) and dipole giant (GDR) resonances from   the differential cross sections at angles $(\theta_{lab}=0.64^{\circ}-13.5^{\circ})$.
The extracted peak strengths of ISGMR exhaust
 almost 100\% of the energy weighted sum rule value in the  nuclei shown.   The average energies of ISGMR are determined to be 
\begin{eqnarray} 
 {\rm E_x(ISGMR)}=\frac{m_1}{m_0}&=&16.6 \pm0.1\rm{ MeV} \,\,\,\rm{ for } \,\,\, ^{90}\rm{Zr, }         \nonumber  \\
 &=&15.4\pm0.1\rm{ MeV} \,\,\,\rm{ for }\,\,\, ^{116}\rm{ Sn},          \nonumber  \\
  &=&13.4\pm0.2\rm{ MeV} \,\,\,\rm{ for }\,\,\,^{208}\rm{ Pb},  
  \end{eqnarray}
  where the $l$-th energy weighted sum rule value is defined  by
   \begin{eqnarray}
   m_l=\int E_x^l\frac{dB(E0,E_x)}{dE_x}dE_x. 
 \end{eqnarray}  
 The strength distribution of  ISDGR have two peak structure in the region of 10$<$E$_x<$30MeV.  An additional  peak is also seen below  E$_x<$10 MeV \cite{1}.

There was an attempt to determine all the values of r.h.s of Eq. (\ref{eq:KA}) from a set of ISGMR energies in several nuclei.  However this attempt got no success since the existing data set was not accurate enough to pin down precisely each value in  Eq. (\ref{eq:KA}).    Another plausible approach to extract the value 
 $K_{\infty}$ from the experimental data is   the framework of self-consistent Hartree-Fock (HF) or Hartree+ 
random phase approximation (RPA) model.  In the self-consistent approach,   a single Hamiltonian, which has good saturation properties,  
is  adopted in all calculations of nuclear matter and finite nuclei so that one can see a direct 
correlation between the incompressibility in nuclear matter and the excitation energy of ISGMR through 
the adopted Hamiltonian.   This approach was quite successful to determine the incompressibility within 
microscopic Skyrme, Gogny and relativistic mean field (RMF)  models \cite{Colo, Khan1,Khan2}. In  ref. \cite{Khan1}, it was also pointed out that  $K_{\infty}$ might not be the best coefficient to fit the energy of ISGMR, but a similar parameter $M_c$ is suggested defined at a density somewhat smaller than the saturation density.

\begin{figure}[htp]
 \includegraphics[clip,bb=0 0 720 540, width=12cm]{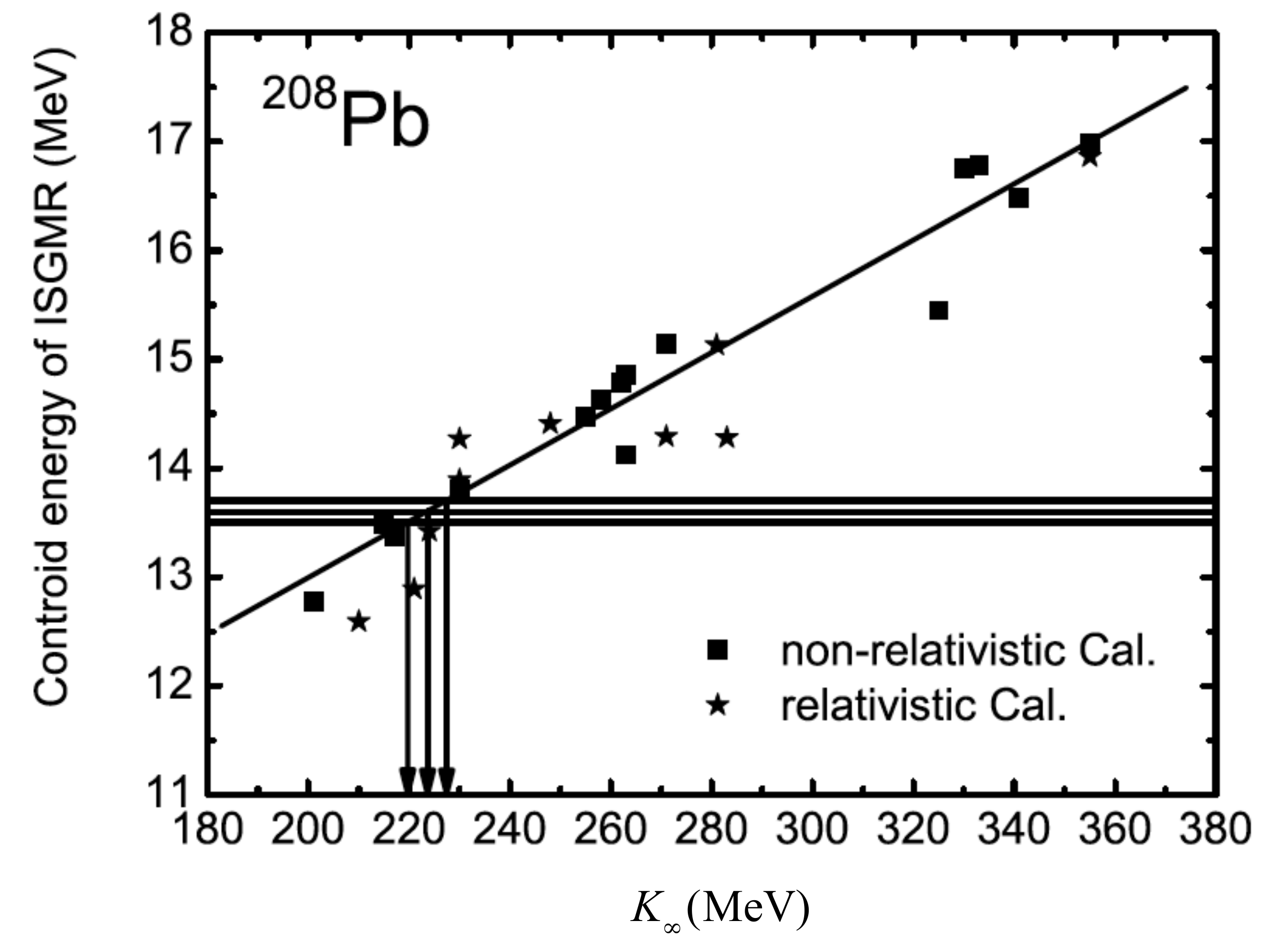}
  \caption{The calculated centroid energies of ISGMR of $^{208}$Pb as a function of $K_{\infty}$. The $K_{\infty}$ is constrained by recent ISGMR experimental data of $(d,d')$ experiments \cite{1} with  the extracted value of $K_{\infty}$ is 223.6 $\pm$ 3.8 MeV.  }  \label{fl}
\end{figure}

In  Fig. \ref{fig:Pb-ISGMR}, the experimental data of ISGMR is compared with the self-consistent HF+RPA calculations with three Skyrme interactions SkP, SLy5 and SkI3 which have the nuclear matter incompressibility  $K_{\infty}$=201, 230 and 258MeV, respectively.  
The empirical strength distributions are better reproduced by SLy5 interaction than the other two interactions. 
  A correlation  between 
 the calculated ISGMR energies  of  $^{208}$Pb with various EDF,  and the nuclear matter incompressibility $K_{\infty}$ is shown in Fig. 3.   Both the excitation energy and $K_{\infty}$ are calculated by using the same EDF.  
 Experimental data are tabulated  in Table 2.  
 We adopt the data of $(d,d')$  experiment $E=13.6 \pm 0.1$MeV from ref. \cite{1} ,which is close to the extracted value from
  $( \alpha, \alpha')$ experiments in refs.  \cite{4,5}.  
 An empirical value of nuclear matter incompressibility is extracted to be $K_{\infty}=223.6\pm3.8 {\rm MeV}$ from this figure. 
However, there are some uncertainty of this value of $K_{\infty}$ which, to some extent, comes from the ambiguity of empirical determination of the ISGMR energy and also from  the theoretical models involved in the microscopic calculations.  Another uncertainty comes from that the mass number dependence of the excitation energies  is not perfectly regular.  Thus the proposed 
empirical incompressibility may depend on how to select the data set of excitation energies of ISGMR.
   Including  the data of superfluid nuclei 
Sn- and Cd-isotopes, the current optimal value of nuclear incompressibility is
\begin{equation}
K_{\infty}=225\pm20 \rm{ MeV}, 
\label{eq:Kexp}
\end{equation}
taking into account  the statistical errors from the experiments and the systematic errors from the theoretical models.


\section{Symmetry energy and terrestrial experiments}
The symmetry energy plays a decisive role to determine the EoS of neutron matter on top of the EoS of symmetric nuclear matter as we  can see in Fig. \ref{fig:EoS}.  From 1990th, tremendous amount of experimental and theoretical efforts have been paid  to explore the symmetry energy at various nuclear matter densities $0<\rho/\rho_0<3$.  At lower density region,  the isovector giant dipole resonances (GDR) give useful information to pin down  the symmetry energy coefficients $J$ and $L$, while the multi-fragmentation process of heavy ion collisions (HIC) provides the empirical information at higher density than the saturation density $\rho_0$.  It was pointed out recently that the mass formula may provide also a  useful information on symmetry energy around the saturation density. 
We will study  the mass formula constrains for the symmetry energy coefficients.  The multi-fragmentation products of heavy ion collisions are  also important to pin down the properties of EoS at higher density than the normal density.  However it is still large uncertainty to extract reliable information of EoS from very complicated multi-fragmentation  results 
 by using  transport models.  Because of this reason, we do not discuss any details of the multi-fragmentation process of heavy ion reactions in  this section.  

\begin{figure}[htp]
\vspace{-1cm}
\includegraphics[clip,width=14cm,
bb=0 0 780 570]{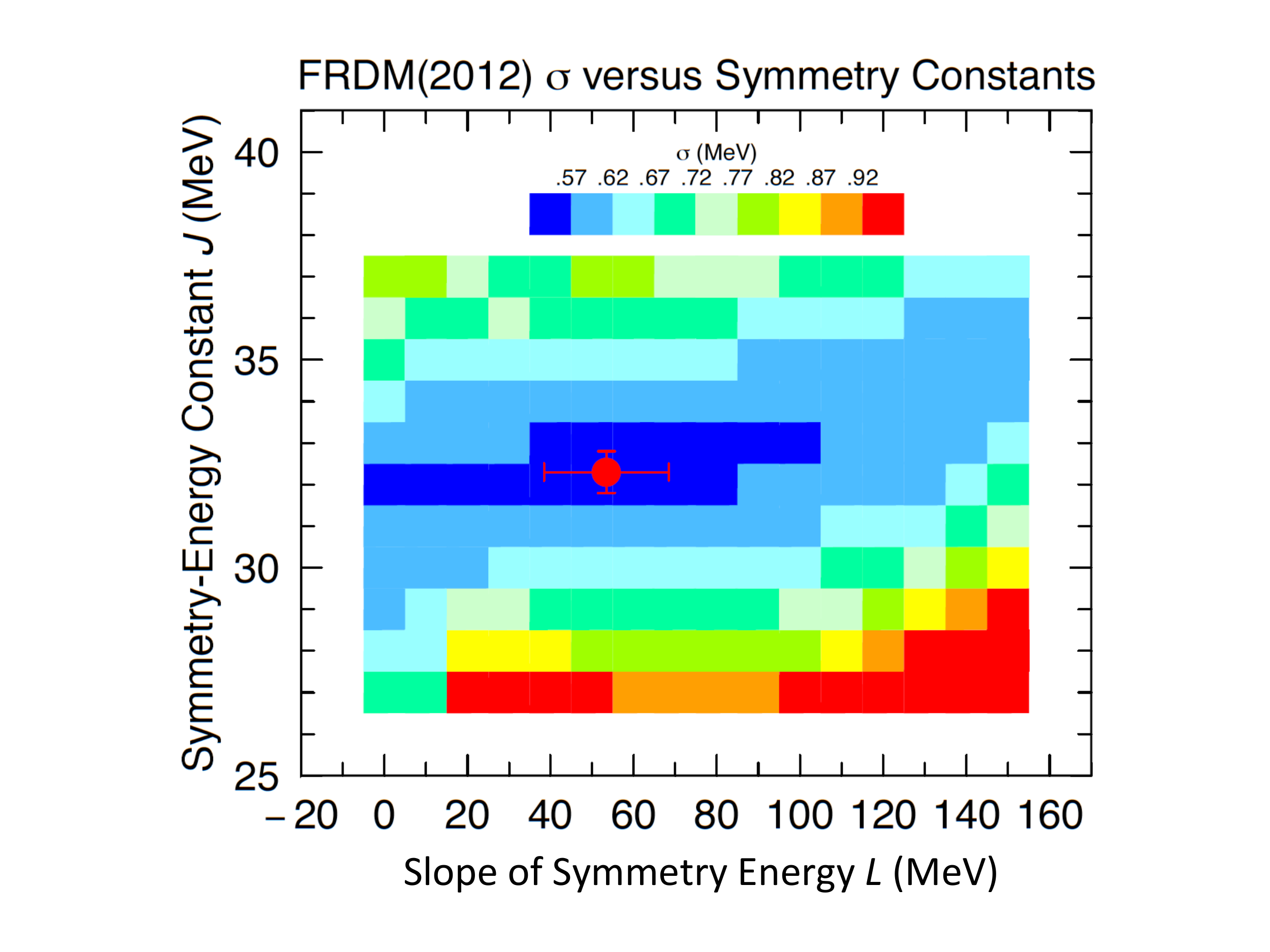}
\caption{Symmetry energy constants $J$ and $L$ versus the mean square deviation $\sigma$ between 
the experimental mass and the FRDM mass formula predictions. Calculated mass model FRDM accuracy are shown by different colors for different values of $J$ and $L$.  The best accuracy  region is indicated by a red dot with uncertainty bars.  This figure is provided by Peter M\"oller and published in 
Atomic Data and Nuclear Data Tables (2016). 
\label{fig:JL-mass}}
\end{figure}

  One of the decisive ingredients of nuclear mass formula is the symmetry energy.  You can see the explicit functional form of mass formula in ref. \cite{Myers,Moller1}. 
    It is curious that how the predicting power of mass formula is sensitive to the symmetry energy coefficients.    
  A recent study of symmetry energy in the mass formula was done by using the finite-range droplet mass model (FRDM) \cite{Moller}.  The FRDM is one of the best mass formulas to predict not only masses of stable nuclei but also unstable nuclei.  
  In the study, the mass parameters  including the symmetry energy coefficients $J$ and $L$ are optimized by using all available experimental data of binding energies for  several thousands nuclei.   
  In Fig.  \ref{fig:JL-mass},  
  the smallest mean square deviation $\sigma$ was obtained by  the optimization process at the values, 
\begin{eqnarray}
J=32.3\pm0.5  \rm{ MeV},  \nonumber \\
L=53.5\pm15  \rm{ MeV } 
\label{eq:JL}
\end{eqnarray}
shown by a red dot with uncertainty bars in Fig. \ref{fig:JL-mass}.   
The values  (\ref{eq:JL})  are consistent with empirical values obtained from GDR, HIC experiments and also from the systematical analysis of excitation energies of isobaric analog states (IAS).

\section{Neutron skin in $^{48}$Ca and symmetry energy}
It has been pointed out that the neutron skin give a useful information to elucidate symmetry energy properties and also neutron matter EoS.  
In previous studies, the doubly magic $^{208}$Pb has been used
as a benchmarking nucleus because the double magicity removes
the effects  which involves additional nuclear structure information such as superfluidity and deformation. 
Many experimental efforts have been  devoted to  determine  the neutron skin $\Delta r_{np}$ of  $^{208}$Pb by measuring  proton elastic scattering \cite{Starodubsky1994b, Zenihiro2010},
coherent pion-photoproduction \cite{Tarbert2014a},
antiprotonic atom X-ray \cite{Kos2007},
and electric dipole polarizability \cite{Tamii2011a}.
 Their
results are in the range of $0.15$-0.21 fm with the error of approximately 0.03 fm. The PREX experiment using parity violating
 electron scattering resulted in $\Delta r_{np}$= 0.33+0.16
-0.18fm\cite{Abrahamyan2012}, which is consistent with other results within very large
statistical error, which prevents precise determination of symmetry energy properties.  

\begin{figure}[htp]
\vspace{-1cm}
\includegraphics[clip,width=12cm,bb=0 0 720 540]{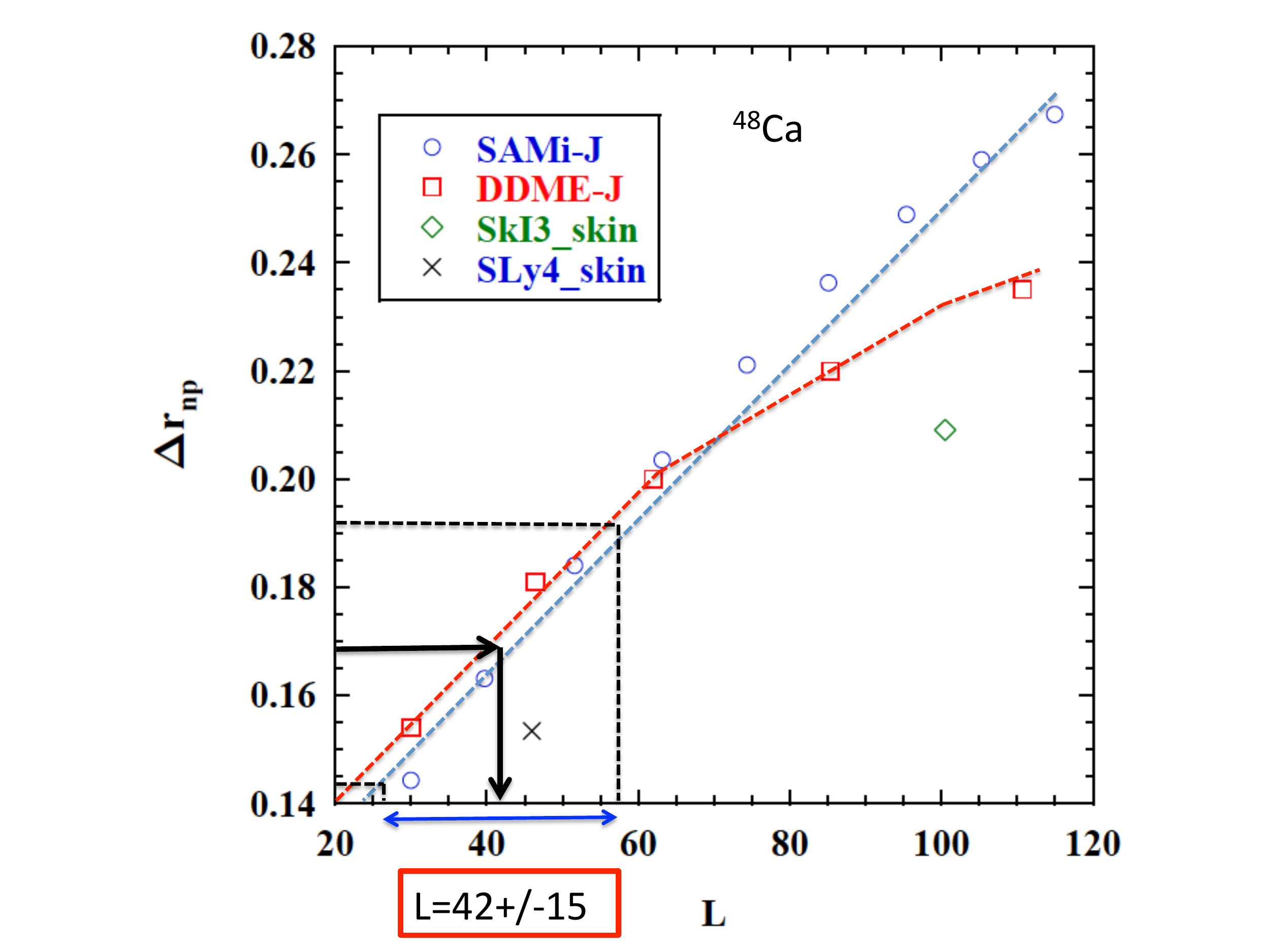}
\caption{A correlation between the neutron skin of $^{48}$Ca and Symmetry energy constants  $L$.
Skyrme EDF SAMi-J and relativistic mean field model DDME-J are adopted together with Skyrme EDF SkI3 and SLy4. 
\label{fig:rnp-L}}
\end{figure}

\begin{figure}[htp]
\begin{centering}
\vspace{-1cm}
\includegraphics[clip,width=12cm,bb=0 0 720 540]{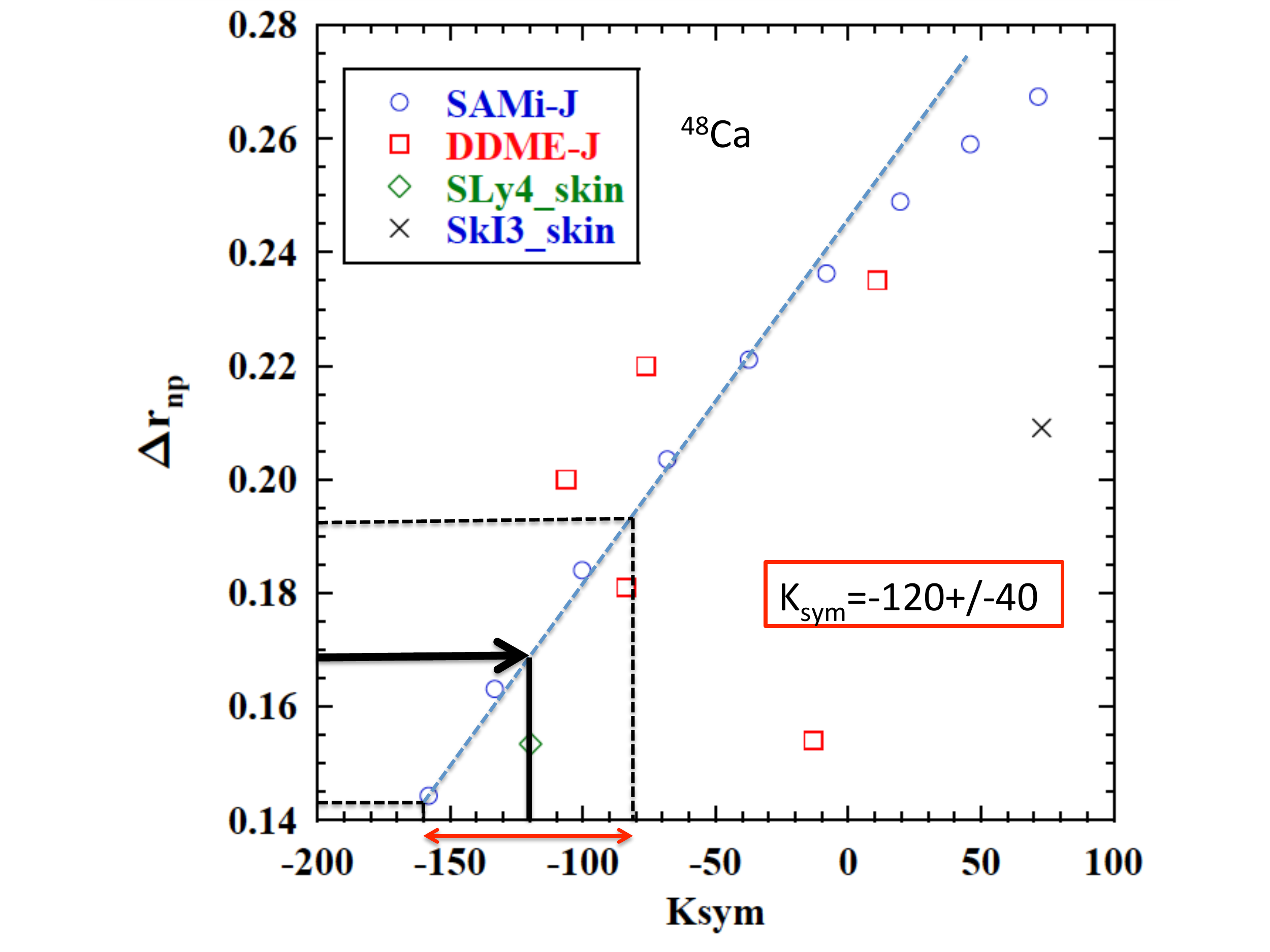}
\caption{A correlation between the neutron skin of $^{48}$Ca and Symmetry energy constants  $K_{sym}$.
Skyrme EDF SAMi-J and relativistic mean field model DDME-J are adopted together with Skyrme EDF SkI3 and SLy4. 
\label{fig:rnp-Ksym}}
\end{centering}
\end{figure}

The accurate measurement of neutron skin of $^{48}$Ca is performed recently by $(p,p')$ experiments \cite{Zenihiro}.
The neutron skin size was determined to be 
\begin{eqnarray}
\Delta r_{np}=0.168+0.025-0.028 \,\,\, \rm{fm} .
\label{eq:dr-np}
\end{eqnarray}
A correlation between the neutron skin $\Delta r_{np}$ of $^{48}$Ca and Symmetry energy constants  $L$ are plotted in
Fig. \ref{fig:rnp-L}  calculated by  
Skyrme EDF SAMi-J and relativistic mean field model DDME-J together with Skyrme EDF SkI3 and SLy4.  
A correlation between the neutron skin of $^{48}$Ca and Symmetry energy constant  $K_{sym}$ are also plotted in
Fig. \ref{fig:rnp-Ksym}  calculated by   the same EDFs as those of  Fig. \ref{fig:rnp-L}.
We can see a clear correlations between $\Delta r_{np}$ and $L$ in Fig. \ref{fig:rnp-L}   with slight model dependence of EDF.  From Skyrme EDF, we can extract the slope parameter $L$ as
\begin{eqnarray}
L=42\pm15  \rm{MeV},  
\label{eq:Ca-L}
\end{eqnarray}
which shows a good agreement with the value extracted from the mass formula FRDM in the previous section.
The correlation between $\Delta r_{np}$  and $K_{sym}$  is  more model dependent of EDF.
Taking Skyrme EDF, we can extract  $K_{sym}$  as
\begin{eqnarray}
K_{sym}=-120\pm40  \rm{MeV}. 
\label{eq:Ca-Ksym}
\end{eqnarray}
For RMF EDF  case, it is difficult to extract the  $K_{sym}$ value since there is no linear correlation between 
$\Delta r_{np}$  and $K_{sym}$.

\section{Summary}
We have critically examined nuclear matter and neutron matter  EoS parameters by using best available terrestrial experimental results.   The nuclear incompressibility  $K_{\infty}$ is extracted 
in comparisons with RPA results of modern relativistic and non-relativistic EDF and systematic  
data of isoscalar giant monopole resonance energy of $^{208}$Pb.  The optimal value is
\begin{equation}
K_{\infty}=225\pm20 \rm{MeV}.    \nonumber 
\end{equation}
The symmetry energy expansion coefficients $J$, $L$ and $K_{sym}$ are examined by recent FRDM mass model and the neutron skin of 
$^{48}$Ca extracted from   $(p,p')$ experiments.  The obtained values from FRDM mass systematics are  
\begin{eqnarray}
J&=&32.3\pm0.5  \rm{ MeV},  \nonumber \\
L&=&53.5\pm15  \rm{ MeV},   \nonumber 
\end{eqnarray}
while the neutron skin experiment of $(p,p')$ experiment gives
\begin{eqnarray}
L&=&42\pm15  \rm{ MeV},    \nonumber \\
K_{sym}&=&-120\pm40  \rm{ MeV}.   \nonumber 
\end{eqnarray}
To determine $K_{sym}$, the results of RMF calculations are excluded.  
These values are consistent with the results of metamodeling analysis of EoS with Skyrme EDF\cite{Jerome2018}.  
It should be mentioned that RMF and RHF seems to prefer slightly larger symmetry energy coefficients than the adopted ones in the present analysis.

\end{document}